\documentstyle[graphics,twocolumn]{mn}
\newcommand{\mincir}{\raise
-2.truept\hbox{\rlap{\hbox{$\sim$}}\raise5.truept 
\hbox{$<$}\ }}
\newcommand{\magcir}{\raise
-2.truept\hbox{\rlap{\hbox{$\sim$}}\raise5.truept
\hbox{$>$}\ }}
\newcommand{\minmag}{\raise-2.truept\hbox{\rlap{\hbox{$<$}}\raise
6.truept\hbox
{$>$}\ }}

\def\lsim{~\rlap{$<$}{\lower 1.0ex\hbox{$\sim$}}}
\def\bsim{~\rlap{$>$}{\lower 1.0ex\hbox{$\sim$}}}

\def\kms{\ {\rm km\,s^{-1}}}
\def\hmpc{\ {\rm {\it h}^{-1}Mpc}}

\def\hmmpc{\ {\rm {\it h}\,Mpc^{-1}}}

 \def\ln{{\rm ln}}

\def\be{\begin{equation}}
\def\ee{\end{equation}}
\def\pmb#1{\setbox0=\hbox{#1}%
\kern-.025em\copy0\kern-\wd0
\kern.05em\copy0\kern-\wd0
\kern-.025em\raise.0433em\box0}

\def\etal{{\it et al.\ }}

\def\delm{\delta_{\rm m}}
\def\delms{\delta_{\rm m}^{\rm S}}
\def\delmu{\delta_{\rm m}^{\rm U}}
\def\delmsr{\delta_{\rm m}^{\rm SR}}
\def\delmsrr{\delta_{\rm m}^{\rm SRR}}
\def\dell{\delta_{\rm L}}
\def\dells{\delta_{\rm L}^{\rm S}}
\def\del2{\delta_{\rm 2}}
\def\del1{\delta_{\rm 1}}

\def\del{\delta}
\def\coav{\langle\delm \vert\dell\rangle}

\begin{document} 
\title[Tracing the Warm Hot Intergalactic Medium in 
the Nearby Universe] 
{Tracing the Warm Hot Intergalactic Medium in
the Local Universe}
\author[M. Viel {\it et al.}]
{M. Viel $^{1}$, E. Branchini $^{2}$, R. Cen $^3$, J.P. Ostriker
$^{1,3}$, S. Matarrese
$^{4,5}$, \cr P. Mazzotta $^6$ \& B. Tully $^{7}$ 
\\ 
$^1$ Institute of Astronomy, Madingley Road, Cambridge CB3 0HA\\
$^2$ Dipartimento di Fisica, Universit\`a di Roma TRE, Via della Vasca
Navale 84, 00146, Roma, Italy\\
$^3$ Princeton University Observatory, Princeton University, Princeton NJ 08544 \\
$^4$ Dipartimento di Fisica `Galileo Galilei', Universit\`a di Padova, via Marzolo 8, I-35131 Padova, Italy \\
$^5$ INFN, Sezione di Padova, via Marzolo 8, I-35131 Padova, Italy \\
$^6$ Dipartimento  di Fisica Universit\`a di Roma `Tor Vergata' Via
Della Ricerca Scientifica 1, 00133 Roma, Italy \\
$^7$ Institute of Astronomy, University of Hawaii, 2680 Woodlawn Drive,
Honolulu, HI 96822, USA\\}

\maketitle
\begin{abstract}
We present a simple method for tracing the spatial distribution  and  
predicting the physical properties
of the Warm-Hot Intergalactic Medium (WHIM),
from the map of galaxy light in the local universe.  
Under the assumption that biasing is local and monotonic
we map the $\sim 2 \hmpc$ smoothed density field of galaxy light 
into the mass density field from which we infer the spatial
distribution of the WHIM in the local supercluster.
Taking into account 
the scatter in the WHIM density-temperature and density-metallicity
relation, extracted from the $z=0$ outputs of  high-resolution and
large box size hydro-dynamical cosmological simulations,
we are able to quantify the probability of detecting WHIM 
signatures in the form of absorption features in the X-ray spectra,
along arbitrary directions in the sky. 
To illustrate the usefulness of this
semi-analytical method we focus on the WHIM properties
in the  Virgo Cluster region.
\end{abstract}

\begin{keywords}
Cosmology: theory -- intergalactic medium -- large-scale structure of
universe -- quasars: absorption lines
 \end{keywords}
           
\section{Introduction}
\label{sec:intro}       
As pointed out for the first time by Fukugita, Hogan and Peebles (1998)
the census of baryons in the low redshift universe is significantly
below expectations. Indeed, the baryon fraction
detected in the Ly$\alpha$ forest at $z\sim 2$ (e.g. Rauch 1998),
which agrees with the recent WMAP measurements of the Cosmic Microwave 
Background (Bennett \etal 2003), is more than a factor of 2
larger than that associated with stars, galaxies and clusters in the 
universe at $z\sim 0$ (Cen \& Ostriker 1999).

Recent studies based on large box-size, hydro-dynamical simulations have
suggested that a significant fraction of the baryons at $z \sim 0$ are found in a gaseous
form, with a temperature between $10^5$ and $10^7$ K in regions of
moderate overdensities $\delta\sim 10-100$ (Ostriker \& Cen 1996; Perna \& Loeb 1998;
Cen \& Ostriker 1999;  Dav\'e {\it et al.} 2001; Cen {\it et al.} 2001), hence providing a
possible explanation to the missing baryons problem.  This Warm-Hot
Intergalactic Medium (WHIM) is found to be in the form of a network of
filaments, and could be observed in the
spectra of background bright sources in the X-ray and far-UV bands in
the form of absorption lines due to elements in high ionization
state, like OVI, OVII and OVIII (Hellsten {\it et al.} 1998, Tripp
\etal 2000, Cen {\it et al.} 2001, Furlanetto {\it et al.} 2004).

The detection of WHIM, however, is challenging.  To date the most
significant detection ($>3 \sigma$) in the X-ray band has been made in
correspondence of an absorption line in the {\it Chandra} spectrum of
the Blazar Mkn 421 that has been interpreted as due to two interviewing
WHIM OVII systems at $z=0.011$ and $z=0.027$ and with column densities
$\magcir 8\times 10^{14}$ cm$^{-2}$ (Nicastro \etal 2004).  Fang {\it
et al.}  (2002) have reported a detection of an OVIII absorption line
along the sight-line toward PKS 2155-304 with {\it Chandra} at $z>0$
that, however, has not been confirmed by subsequent observations
performed with {\it XMM-Newton}.  The identification of the absorbers
in the very local universe with WHIM structures in the local
group/supercluster is even more controversial (Nicastro \etal 2003,
Savage \etal 2003).  Finally, emission by the WHIM could also be
important and some detections have been claimed, either by observing
X-ray excess in clusters of galaxies (e.g. Kaastra \etal 2003) or by
observing soft X-ray structures associated with galaxy overdensities
(Zappacosta \etal 2004).

Although these observations are very encouraging and 
seem to indicate that we have just started to detect the WHIM,
various problems are still preventing us from a detailed investigation of its properties.

First of all, the detection of WHIM is particularly difficult and
requires very sensitive UV and X-ray detectors, both for absorption and
for emission processes.  With the presently available instruments, the
WHIM absorption lines can only be detected in very high signal-to-noise
spectra either obtained with very long exposures of bright
extragalactic sources or by observing variable sources like
Blazars when they undergo an
outburst phase.  In either case, expected detections will be so rare
that a detailed study of the WHIM will have to wait for future
observations by next generation X-ray satellites such as {\it XEUS} and
{\it Constellation-X} (Chen {\it et al.} 2003, Viel {\it et al.} 2003).

The second issue, on which we will focus in this work, is the
difficulty in modeling the WHIM. Hydro-dynamical simulations of
structure formation in the context of cold dark matter models have
certainly provided a major contribution to the understanding of the
physics of the WHIM and to the characterizing its spatial
distribution. However, in general, they are not designed to model the
actual WHIM distribution in our universe, i.e. they do not tell us
where we should look for the WHIM in our cosmic neighborhood.  The
hydro-dynamical simulations of Kravtsov {\it et al.} (2002) and, more
recently of Yoshikawa {\it et al.} (2004), constitute the only attempt
to model the gas distribution in the Local Supercluster region to
address WHIM detectability.  The simulations by Kravtsov {\it et al.}
(2002) make self-consistent use of constraints from the MARK III
catalog of galaxy peculiar velocity to reproduce the large-scale mass
density field within $\sim 100 \hmpc$, including major nearby
structures like the Local Group, the Local Supercluster, Virgo and Coma
clusters (Klypin \etal 2003).  However, their peculiar velocity
constraints are only effective above a (Gaussian) resolution scale $> 5
\hmpc$ which is significantly larger than the typical widths of the
WHIM structures (Viel \etal 2003).  The constraints of Yoshikawa {\it
et al.} (2004) are derived from the redshift space positions of IRAS
1.2Jy galaxies and are effective on a $5 \hmpc$ Gaussian scale.  As a
consequence, in both models the actual position of the WHIM absorber
within the resolution element of a single numerical simulation is
essentially random.  A large number of constrained hydro-dynamical
simulations, characterized by independent sets of Fourier modes on
sub-resolution scales, would then be required to produce independent
realizations of the intergalactic gas in the local universe from which
a {\it probability} of detecting a WHIM absorber at a given spatial
location can be computed.

Unfortunately, this approach is too time-consuming to be
successfully implemented using hydro-simulations and to allow
an accurate study of all the possible physical processes
which can influence the properties of the WHIM at $z\sim 0$ such as the effect of
different amount of feedback in the form of
galactic winds, the distribution of metals 
and the effect of having different temperature-density relations.

For this reason we have developed an alternative semi-analytical modeling,
still based on the same probabilistic approach, 
in which we use the existing  results of hydro-dynamical simulations
at $z=0$ to  trace the intergalactic gas 
from the spatial distribution of galaxy light 
in the local supercluster.
The intrinsic scatter in the gas vs. galaxy light relation
is accounted for by performing independent realizations of 
the gas distribution in the local universe which we can produce by 
means of fast Montecarlo techniques.
Another advantage of our method is that our constraints, that derive
from mapping the galaxy light, apply down to scales of
 $\sim 1  \hmpc$, comparable to the width of WHIM structures and much
smaller than the galaxy correlation length, hence allowing us 
to trace the local WHIM with 
higher resolution.
Obviously, these considerations are valid as long as 
light to gas  mapping is local and well modeled by our
reconstruction technique, which we will show is indeed the case.

The outline of the paper is as follows.
In Section 2 we will describe the galaxy 
catalog used in
our analysis. Section 3 presents the hydro-dynamical cosmological
simulations used to model the WHIM structures.
In Section 4 we will describe the method to produce X-ray absorption
spectra along an arbitrary direction in the sky starting from the
galaxy luminosity map in the local universe and from the physical
properties of WHIM structures as extracted from the hydro-dynamical
simulations. 
Section 5 contains a testing of the modeling, while an
application to the local universe is presented in Section 6. Section 7
contains our conclusion and the future developments of the method proposed.

\section{The Nearby Galaxy Luminosity Field}
\label{sec:tullycat}       
\begin{figure*}
\center
\resizebox{1.0\textwidth}{!}{\includegraphics{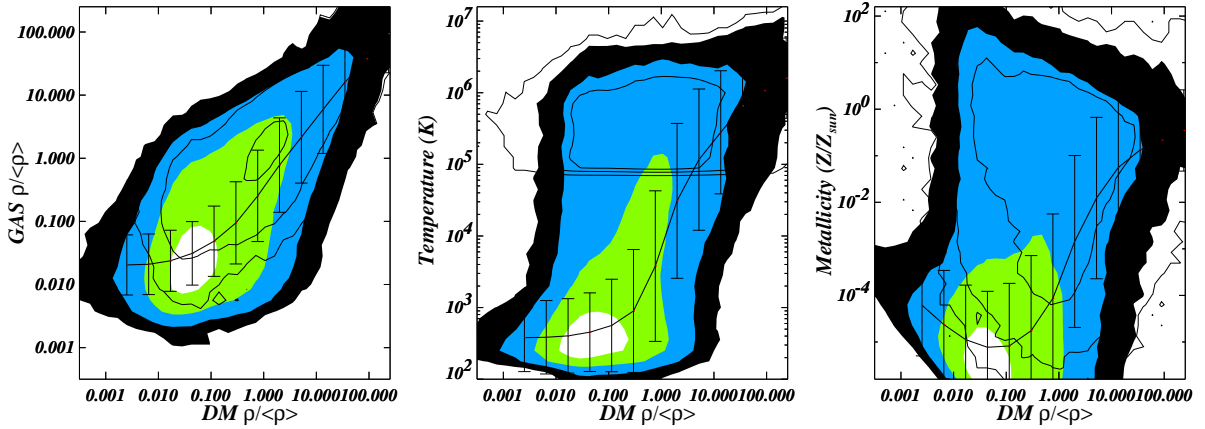}} 
\caption{{\it Left panel:} Gas density as a function of mass density (in units of the
mean density) as extracted from the $z=0$ output of a $\Lambda$CDM
simulation with a box size of $25 \hmpc$ (our fiducial simulation, see Section \ref{hydro}). 
The line represents the median value in each bin and error bars
are the $1\,\sigma$ values. Filled contour plots represent the number density
of points in the simulation  (number density  increases by an order of
magnitude at each contour level). The empty contours show the number
density of those pixels with $T>10^5$ K (number density increases by an order of
magnitude at each contour level). 
{\it Central panel:} Gas temperature as a function of
mass density.
{\it Right panel:} Gas metallicity (in units of solar metallicity) as a
function of mass density.}
\label{fig1}
\end{figure*}

The galaxy sample considered in this work has been extracted from the
latest version of the Nearby Galaxies Catalog (Tully 1988).  The
extraction is a cube with cardinal axes in supergalactic coordinates
that extend to $\pm 1500$~km~s$^{-1}$, centered on our Galaxy.
This boundary fully encompasses the Virgo, Ursa Major, Coma~I, and
Fornax clusters.  The database strives for completeness among all galaxy
types known.  The all-sky optical surveys accumulated in ZCAT
(http://cfa-www.harvard.edu/$\sim$huchra/zcat/zcom.htm)
provides completeness at all but low galactic latitudes for high surface
brightness galaxies.  Though dated, the HI survey of Fisher and Tully
(1981) provides the most complete all-sky coverage of low surface
brightness irregular galaxies.   The catalog lists angular positions,
redshifts, and B-band luminosities.   There has been compensation for
the small incompleteness associated with the zone of avoidance through
the addition of fake sources through reflection of real sources at
higher latitudes.  In total, the catalog contains 1968 real sources
and 106 fake sources within the cube of diameter 3000~km~s$^{-1}$.
There is correction for missing information with distance but it is
very small.  The information to be used is the {\it luminosity
density} at B-band (not the galaxy number density).  The sample has a
faint luminosity clip at $M_B = -16 +5{\rm log~H}_0/100$.  The catalog
is essentially complete for galaxies of this luminosity at the
boundaries of the cube along the cardinal axes.  Corrections for lost
light amount to 30\% at the far corners of the cube.

The analysis to be discussed is performed in real rather than redshift
space.   The finger of god velocities in clusters have been collapsed
so the clusters have depths comparable to their projected dimensions.
Then distances have been assigned to groups and individual galaxies in
accordance with the output of a Least Action Model (Peebles 1989;
Shaya, Peebles, and Tully 1995) constrained by 900 individual distance
measurements.   This model provides a reasonable description of galaxy
flows in the neighborhood of the Virgo Cluster.

\section{Hydro-dynamical simulations}
\label{hydro}

\begin{figure*}
\center
\resizebox{1.0\textwidth}{!}{\includegraphics{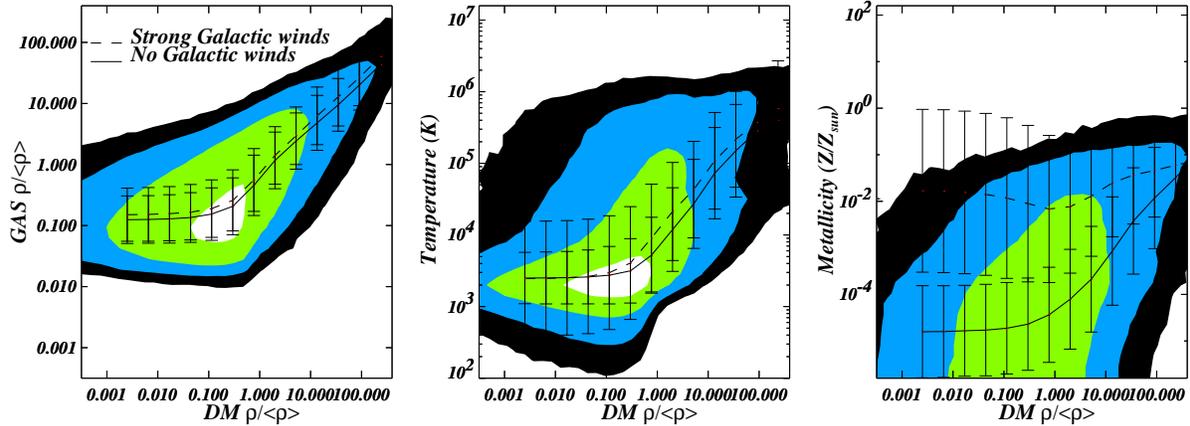}} 
\caption{{\it Left panel:} Gas density as a function of mass density (in units of the
mean density) as extracted from the $z=0$ output of a $\Lambda$CDM
simulation of $11 \hmpc$ (see Section \ref{hydro} and Cen, Nagamine,
Ostriker 2004 for more details). 
The lines represent the median value in each bin and error bars
are the $1\,\sigma$ values. The continuous lines in all the panels refer to a simulation
without galactic winds while the dashed ones are for a simulation with
strong galactic superwinds.
Filled contour plots represent the number density
of points in the simulation  without galactic winds (number density increases by an order of
magnitude at each contour level). 
{\it Central panel:} Gas temperature as a function of
mass density.
{\it Right panel:}  Gas metallicity (in units of solar metallicity) as a
function of mass density.}
\label{fig2}
\end{figure*}

The fiducial hydro-dynamical simulation used here is a $\Lambda$CDM model with
the following parameters: $H_0=100h$ km s$^{-1}$Mpc$^{-1}$ with $h=0.67$,
$\Omega_{0m}=0.30$, $\Omega_{0b}=0.035$, $\Omega_{\Lambda}=0.70$, $\sigma_8=0.90$,
and the spectral index of the primordial power spectrum $n=1$. The
box-size of the simulation is $25 \hmpc$ comoving  on a uniform
mesh with $768^3$ cells. The comoving cell size is 32.6 $h^{-1}$ kpc
and the mass of each dark matter particle is $\sim 2 \times 10^7$
M$_{\odot}$ (further details can be found in Cen {\it et al.} 2003).
The simulation includes galaxy and star formation, energy feedback
from supernova explosions, ionization radiation from massive stars and
metal recycling due to SNe/galactic winds. Metals are ejected into the
local gas cells where stellar particles are located using a yield
$Y=0.02$ and are followed as a separate variable adopting the standard
solar composition. This simulation can be considered a  good
compromise between box size and resolution to investigate the WHIM
properties (see discussion in  Cen {\it et al.} 2001).

The X-ray and UV background of the simulation at $z=0$, 
which are needed to properly simulate X-ray absorptions,
are computed self-consistently from the sources in the 
simulation box.  
We have checked that there are very small differences between the
hydro-simulation background and the estimate from Shull {\it et al.}
(1999): $I_{UV} = I_{UV}^0 (E/13.6\rm{eV})^{-1.8}$, with
$I_{UV}^{0}=2.3 \times 10^{-23}$ erg cm$^{-2}$ Hz$^{-1}$ sr$^{-1}$
s$^{-1}$ for the UV background (which includes a contribution from AGNs
and starburst galaxies), and
$I_{X}=I_{X}^{0}(E/E_{X})^{-1.29}\exp{(-E/E_{X}) }$, with
$I_{X}^{0}=1.75 \times 10^{-26}$ erg cm$^{-2}$ Hz$^{-1}$ sr$^{-1}$
s$^{-1}$ and $E_{X}=40$ keV (Boldt 1987; Fabian \& Barcons 1992;
Hellsten {\it et al.} 1998; Chen {\it et al.} 2003), for the X-ray
background.  The difference between the amplitude of the UV and X-ray
background of the hydro-simulations and the estimates above is less
than a factor of 2, in the range 0.003 \mincir E (keV) \mincir 4.  
Among the two backgrounds, the X-ray one plays a major role in
determining the statistics of the simulated spectra.  We will assume
that the amplitude of the ionizing backgrounds scales like $(1+z)^3$,
in the redshift range $0<z<1$ (this is in agreement with the evolution
of the radiation field in the simulation). This assumption will not influence
significantly our results: given the fact that we will generate
mock-spectra at $z \sim 0$.

Along with the ionizing background, the other 
quantities extracted from the hydro simulations which are relevant for our work 
are the mass density, and the  density,  temperature and 
metallicity of the gas. All these quantities have been re-binned
on a coarser $256^3$ cubic grid, with cells of 
comoving size  97.6 $h^{-1}$ kpc. In the rest of the paper we will
label as {\it fiducial} the simulation with a box of 25 $h^{-1}$
comoving Mpc and $768^3$ cells rebinned into a mesh of $256^3$. 
In the following we will refer to these quantities
as the {\it unsmoothed} fields and will indicate them with the upper index
${\rm U}$. 

The gas density-mass density relation in the hydro-dynamical simulation
is shown  by the contour plot in the left panel of Figure
\ref{fig1}, where we also report the median value in each density bin
(continuous line) and the rms values.  Similarly, the central and right panels 
show the gas temperature and gas metallicity as a function of  mass density,
The filled contour plots represent, in all panels, the number density of
mass density elements in a given bin of gas-density/mass-density, 
gas-temperature/mass-density and gas-metallicity/mass-density
in the hydro-dynamical simulation. The empty contours represent the
number density of gas elements with $T>10^5 $K, i.e. the gas
effectively responsible for absorption. 
We prefer to show these plots
instead of the usual gas density-temperature and gas
density-metallicity relation (as in Viel {\it et al.} 2003), 
because our modeling starts from the mass distribution and uses
these relations to predict the WHIM distribution and properties.

The relations shown in Figure \ref{fig1}  are in reasonable agreement with the recent
results obtained by Yoshida {\it et al.} (2002 - their Figure 2) and
also by Dav\'e {\it et al.} (1999), who studied the gas cooling processes
in the framework of Smoothed Particle Hydro-dynamics (SPH)
simulations. We note that the scatter in temperature can be very high
for values of $\log(1+\delmu)>-1$, spanning more than one order of
magnitude.  The mass density-metallicity relation in the right
panel is as expected: low density regions are the most
metal poor, while at large densities the metallicity can be solar.  In
this plot the scatter is extremely large for the intermediate density
regime, i.e. $-1<\log(1+\delmu)<0.5$, while it is significantly
smaller for larger overdensities.  This plot has to be compared with
similar plots shown in Cen \& Ostriker (1999b), but here we find that
the scatter is somewhat larger than in previous simulations. This can
be due to the different amount of feedback in the hydro-simulation. The
effect of feedback on WHIM structures will be investigated in a future work.

However, to have a first hint on what properties of the WHIM are
sensitive to the physics implemented into the hydro-dynamical
simulations we have analyzed the $z=0$ of two other simulation with a
smaller box size and at higher resolution. These simulations have a box
size of $11 \hmpc$ comoving on a uniform mesh of 432$^3$ elements and
with the following cosmological parameters:
$\Omega_{0m}=0.29,\Omega_{0b}=0.047, \Omega_{\Lambda}=0.71,
\sigma_8=0.85, n= 1, h=0.7$, and are extensively described in Cen et
al. (2003) and Cen, Nagamine \& Ostriker (2004). The main difference
between this second set of simulations and our fiducial one relies in
the presence of galactic superwinds. These are generated as feedback
from star formation and are normalized to match Lyman Break galaxies
observations. In Figure \ref{fig2} we plot the
gas density, the gas temperature and the gas metallicity vs. dark
matter density relations as in Figure \ref{fig1}. We note that, while
the gas density-dark matter density relation is very similar, the
temperature plot and the metallicity one show some differences. In
particular, it seems that at a fixed density the temperature of the gas is about a factor 
two smaller
 than in the larger box size one. This can be
determined both by the different feedback implementations and by the
smaller box size which samples a volume which is certainly not large enough
to contain massive structures.  Another major difference is in the
metallicity-dark matter density diagram, where one can see that in the
case of strong galactic winds, represented by the dashed line, the low
density regions are significantly metal enriched. For the higher
density regions with $10<\delta<100$ the discrepancy is smaller but
still it seems that in the presence of galactic winds their metallicity
can be a factor $\mincir 2$ higher than in the no winds case.  This is
a very similar trend to that found by Cen, Nagamine \& Ostriker (2004)
at higher redshift in the Lyman$-\alpha$ forest and shows that winds
are effective in polluting the low density intergalactic medium with
metals. Another feature which is worth stressing is the fact that both
the temperature-dark matter density and the metallicity-dark matter
density relations in the simulation with winds show  larger scatter
(about a factor two) when compared to the simulations without winds.
The results with no winds (continuous line)  are
however in reasonable good agreement with our fiducial simulation ( 
Figure \ref{fig1}), which contains a different and probably
less effective implementation of feedback by galactic winds.

In this Section we have analyzed the properties of the WHIM as derived
from hydro-dynamical simulations with different box-sizes and different
implementation of physical processes. Among all the effects we found
that the presence of galactic winds and the box sizes are important in
determining the properties of the WHIM in particular its temperature
and metallicity.  In the semi-analytic modeling we are going to
propose in the following Sections, which is of course less accurate than
the hydro-dynamical simulations described here, an implementation of the different
relations shown in Figures \ref{fig1} and \ref{fig2} will be
straightforward, allowing us to produce many realizations of the local
universe WHIM distribution.

\section{From Galaxy Light to WHIM}
\label{sec:mapping}

In this Section we describe a method to 
infer the spatial distribution of the Intergalactic Medium  from the 
observed galaxy distribution in the local universe. The method 
consists of four steps, described in the next 
sub-sections, through which the 
B-band luminosity density field obtained from 
the Tully catalog is used to infer the spatial distribution 
of the dark matter. Then, from the dark matter density field we predict  the 
spatial distribution, temperature and metallicity of 
the intergalactic gas in the local universe.

\subsection{Light to Smoothed Mass Mapping}
\label{sec:lightmapping}       

To map the observed light distribution into the mass distribution we have 
applied the inversion method of Sigad, Branchini and Dekel (2000). 
As pointed out by Dekel and Lahav (1999)
the relation between two 
random fields like $\dell$ and  $\delm$,
can be described by the conditional probability distribution 
$P(\delm | \dell)$.
If this relation  were deterministic, then
the relation between the two fields
would be completely specified by the mean biasing function
\be
\coav =\int P(\delm | \dell) \delm d\delm \; .
\label{eq:cond_def}
\ee
Under the hypothesis that the relation between $\dell$ and $\delm$ 
is  deterministic and monotonic 
then the mean relation $\coav$ can be obtained
by equating the cumulative distribution of  $\dell$,  $C_{\rm L}(\dell)$,
and $\delm$,  $C_{\rm m}(\delm)$, at given percentiles:
\be
\coav = C^{-1}_{\rm m} [C_{\rm L}(\dell)] \ ,
\label{eq:cc}
\ee
where $C^{-1}_{\rm m}$is the inverse function of $C_{\rm m}$.

In general, however, the relation is not deterministic since
the value of the  $\dell$ at a generic location might 
not be solely determined by $\delm$. 
In fact, other physical quantities, such as the gas temperature, or for
example 
local shear,  affect the process of galaxy formation
and evolution (e.g. Blanton \etal 1999)
and thus can introduce a scatter in the 
relation usually referred to as stochasticity. 
In presence of stochasticity, $C_{\rm m}^{-1} [C_{\rm L}(\dell)]$ may still be a
good approximation for $\coav$ as long as the biasing is monotonic.
Branchini \etal (2004) applied this method to
the simulation of Cen {\it et al.} (2001) to test the validity
of this approach. More precisely, they have estimated the mean biasing
relation from the cumulative distribution function of the $\sim 2 \hmpc$-smoothed
density fields of galaxy light and mass in the hydrodynamical simulation.
They found that the $\coav$ measured in the simulation
agrees well with its estimate, $C_{\rm m}^{-1} [C_{\rm L}(\dell)]$,
within the errors.

The method has been implemented as follows.  First we have considered
the luminosity overdensity field at the generic grid point position,
$\dells(\bf{x})$, and have assigned a mass overdensity $\delms(\bf{x})$
$=C_{\rm m}^{-1} [C_{\rm L}(\dells(\bf{x}))]$.  The result is a mean
mass overdensity field that we have perturbed by adding a Gaussian
scatter that accounts for both the stochasticity and the uncertainties
affecting the determination of $\coav$. 
A detailed study of the mean biasing function of the B-band galaxy light
density of the Tully catalog will be presented by Branchini \etal (2004).
Here, we only stress that the resulting biasing function is 
monotonic, i.e. it satisfies the main requirement of the light to mass mapping 
procedure, and is in good agreement with the mean B-band light biasing function
measured in the  hydrodynamical  simulation of Cen {\it et al.} (2001), 
once that sampling errors and cosmic scatter are accounted for.

Using the mean biasing function measured by Branchini \etal (2004)
and an estimate of the scatter around the mean obtained from the 
hydrodynamical simulation, we have applied our mapping procedure 
to the observed B-band galaxy light map and have  
obtained 30 different realizations of the mass density field in the
local universe that will constitute one of the input of the 
reconstruction procedure presented in this work.

\begin{figure*}
\center
\resizebox{0.95\textwidth}{!}{\includegraphics{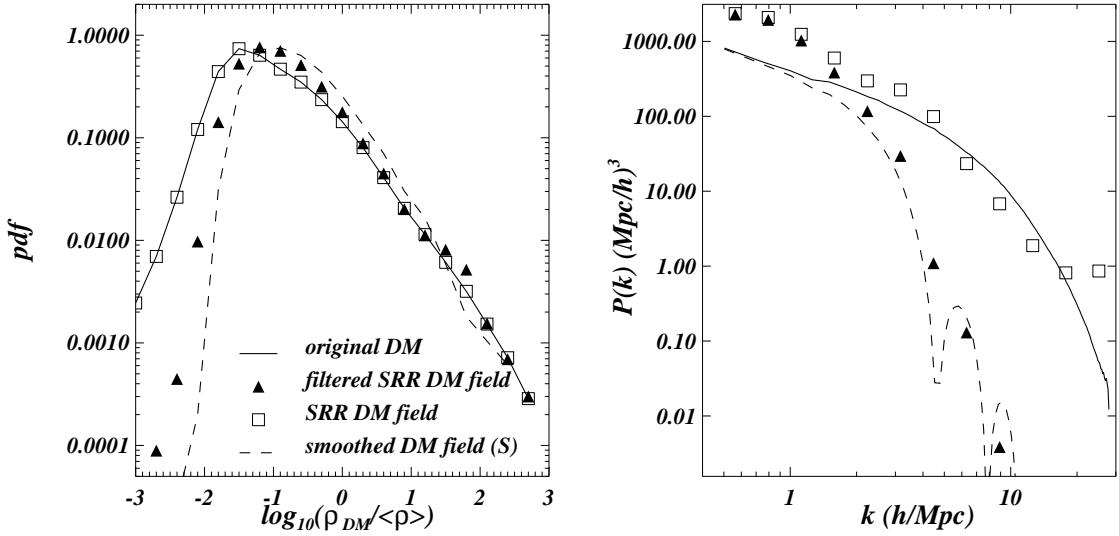}} 
\caption{{\it Left:} The one point probability distribution function  
for the original lognormal density field 
extracted from
hydro-dynamical simulation $\delta_{\rm LN}$
(continuous line); 
for the same field smoothed with a Top Hat filter of
radius $2500  \kms$ (dashed line);
for the reconstructed density field, $\delmsrr$ (open squares)
and for the reconstructed field smoothed on the same scale
(filled triangles).
{\it Right:} Power spectra of the same four density fields, characterized by
the same symbols/line-styles.}
\label{fig3}
\end{figure*}

\subsection{Smoothed Mass to Unsmoothed Mass Mapping}
\label{sec:srr}       

The mass field  $\delms$ obtained from the light distribution
is smoothed on a Gaussian scale $r_s=95 \kms$ which is too large
to investigate the properties of the X-ray absorbers
in the local universe.

The problem of how to infer an unsmoothed mass density field,  $\delmu$,
from a smoothed one, i.e. how to recover the correlation properties
on sub-resolution scale from large scales constraints, 
is well known. A self-consistent solution is found when $\delmu$ is
 Gaussian and the power spectrum of mass density 
fluctuations is known {\it a priori} (Hoffman \& Ribak 1991).
Unfortunately, in our case the nonlinear evolution has caused both
$\delmu$ and  $\delms$ fields to deviate from Gaussianity
and we cannot use conventional reconstruction techniques.

Therefore, we have developed an alternative reconstruction strategy that only assumes 
{\it a priori} knowledge of the mass power spectrum which we set to be the same
as in the hydro-simulation discussed in Section 3.
This method consists of two steps:

${\it i)}$ We start from any of the 30 realizations of $\delms$ and
increase the power on sub-resolution scales by adding Fourier modes
with random phases for wavenumbers larger than $k_S=2\pi/r_S$ in such a
way to follow the power spectrum of the hydro simulations and to
preserve the original spectral shape and amplitude on wavenumbers
larger than $k_S$.  Motivated by the fact that the one point
probability distribution function [PDF] of $\delmu$ should be close to
lognormal we add the power on the lognormal field $\delta_{\rm LN}^{\rm
U} \equiv \ln(1 + \delmu)/<\ln(1 + \delmu)>-1$ rather than to $\delmu$.
We label the new mass density field as $\delmsr$ to indicate that
random power has been added to the smoothed field on sub-resolution
scales.

${\it ii)}$  At this point, we enforce the correct mass PDF
by performing  rank ordering between 
$\delmsr$ and the mass density field extracted from the 
hydro-simulation, $\delmu$. We indicate the resulting mass
density field as $\delmsrr$.

The whole purpose of this SRR reconstruction
procedure is to obtain an unsmoothed density 
field, 
$\delmsrr$, with a power spectrum that resembles that
of the original  hydro-dynamical simulations over the whole range of wavenumbers
and possesses the same PDF of $\delmu$.
It is worth stressing that this procedure is not self-consistent
and therefore we do not expect to reconstruct the original mass 
density field on a point-by-point basis.
However, as we will show in the next section, 
this reconstruction method  works reasonably well in regions of enhanced
 density  where the warm hot phase of the intergalactic gas typically resides.

\subsection{Mass to Intergalactic Gas Mapping }
\label{sec:igm}       

As a further step we map the mass overdensity field, $\delmsrr$,
into gas overdensity, temperature and metallicity.
We do that by using the relations between these quantities 
determined
from the hydro simulation described in Section 2 and
displayed in Figure \ref{fig1}.
The large scatter in these plots is highly non Gaussian.
As a consequence
we cannot assign any of the gas properties as a Gaussian deviate
about the mean value of the gas density, temperature and metallicity
measured in each mass density bin.
Instead, we take the number of elements in each bin of the contour plots
shown in Figure \ref{fig1} to be proportional to the joint probability
of any two quantities, for example, $(\delta_{\rm gas},\delmu)$, and assign the gas 
properties ($\delta_{\rm gas}, {\rm T}_{\rm gas}$ and ${\rm Z}_{\rm gas}$)
to the generic mass element by Montecarlo sampling each of the  
joint probability functions $P(\delta_{\rm gas},\delmu)$, 
$P({\rm T}_{\rm gas},\delmu)$, and  $P({\rm Z}_{\rm gas},\delmu)$,.

This allows us to infer the spatial distribution of the Intergalactic
Medium (IGM) in the
local universe that is guaranteed to reproduce the correct one point
PDF of the density, temperature and metallicity of the gas in the
hydro-simulation.  We stress that in the simple modeling presented
here an implementation of different metallicity-temperature relations,
motivated by physical arguments, will be straightforward.

 It is worth noting that Montecarlo sampling of the probability
distribution function is an approximate procedure that has the
advantage of being easy to implement. However, the large scatter introduced by
this procedure, that reflect the one present in the probability functions,
may spoil our ability to reconstruct the properties 
of the X-ray absorbers. We will check below if this is indeed the case.

\subsection{Intergalactic Gas to X-ray Absorbers Mapping}
\label{sec:xabsorbers}

\begin{figure*}
\center
\resizebox{1.0\textwidth}{!}{\includegraphics{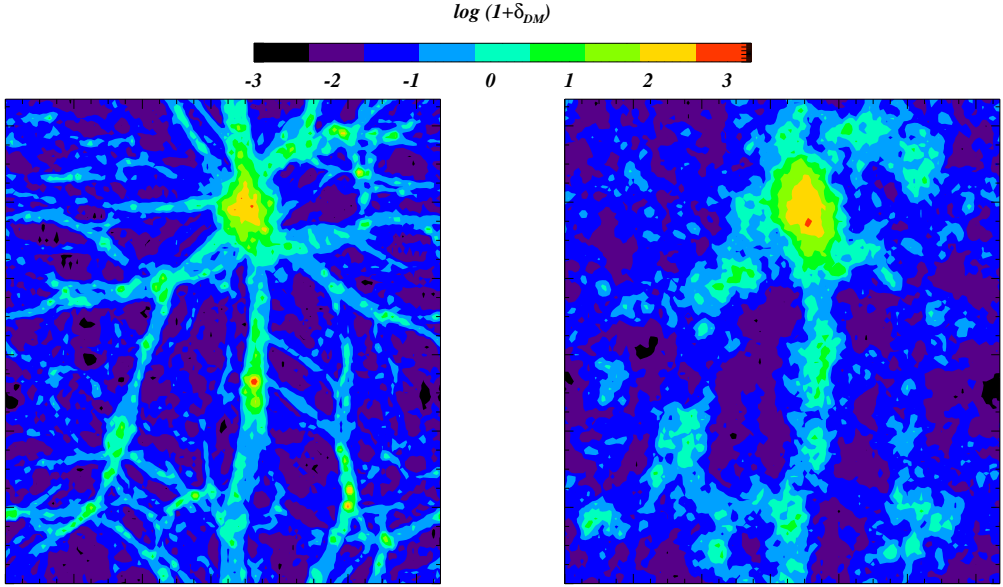}} 
\caption{{\it Left :} The mass density field on a
slice of $25 \times 25 \times 0.2 \hmpc$  around the most massive object found in the
hydro-dynamical simulation. {\it Right:} The reconstructed dark
matter density field $\delmsrr$ in the same region. 
The most massive structures at the
typical overdensities of WHIM are preserved.}
\label{fig4}
\end{figure*}

After having simulated the gas distribution, the photoionization code
CLOUDY (Ferland {\it et al.} 1998) is used to compute the ionization
states of metals, taking into account  the UV and X-ray 
background.
The density of each ion is obtained through: $n_I(x)= n_H(x) X_I
Y_{Z_{\odot}} Z/Z_{\odot}$, where $X_I$ is the ionization fraction of
the ion as determined by CLOUDY and depends on gas temperature, gas
density and ionizing background, $n_H$ is the density of hydrogen
atoms, Z is the metallicity of the element and $Y_{Z_{\odot}}$ is the
solar abundance of the element.  

Among the heavy elements that can produce absorption lines in the
X-rays Oxygen is the most abundant one and produces the strongest
lines.  In this work we only simulate
absorption spectra for the two strongest transitions of the ions OVII
($E=0.57$ keV, $\lambda=21.6 \AA$, $f=0.7$) and OVIII ($E=0.65$ keV,
$\lambda=18.97\AA$, $f=0.42$), with $f$ the oscillator strength.
As pointed out by Viel \etal (2003), in the temperature and density 
range typical of WHIM, these two ions have a  recombination time 
smaller that the Hubble time and thus the approximation of
photoionization equilibrium should be reasonable.

Given the density of a given ion along the line-of-sight, the optical depth 
in redshift-space at
velocity $u$ (in km s$^{-1}$) is 
\be \tau_I(u)= {\sigma_{0,I}c\over H(z)} \int_{-\infty}^{\infty}dy\,
n_{\rm I}(y) {\cal
V}\left[u-y-v_{\parallel}^{\rm I}(y),b(y)\right]
\label{tau} 
\ee 
where $\sigma_{0,I}$ is the cross-section for the resonant absorption
and depends on $\lambda_I$ and $f_I$, $y$ is the real-space coordinate
(in km s$^{-1}$), ${\cal V}$ is the standard Voigt profile normalized
in real-space, $b=(2k_BT/m_I c^2)^{1/2}$ is the thermal width and we
assume that $v_I=v_{IGM}$ is the peculiar velocity. 
Velocity $v$ and redshift $z$ are related
through $d\lambda/\lambda=dv/c$, where $\lambda=\lambda_I(1+z)$. For
the low column-density systems considered here, the Voigt profile is
well approximated by a Gaussian: ${\cal V}=(\sqrt{\pi}
b)^{-1}\exp[-(u-y-v_{\parallel}^{\rm I}(y))^2/b^2]$.  The X-ray
optical depth $\tau$ will be the sum of the source continuum, which we
assume we can determine, and that of eq. (\ref{tau}). Finally, the
transmitted flux is simply ${\cal F}=\exp(-\tau)$.

\section{Testing the modelling}
\label{method}

In this Section we use the hydro simulation described
in Section 3 to test 
the ability of our reconstruction procedure to recover
the correct spatial distribution and physical properties
of the WHIM and the OVII and OVIII absorbers associated to it.

The goodness of the first step of our procedure, i.e. the ability
of tracing the smoothed mass distribution using galaxy light,
has been extensively tested by Branchini \etal (2004).
The authors have 
shown that the mapping procedure is robust and
unbiased and that random errors in 
the light-to-mass mapping are much smaller than the stochasticity 
in the biasing relation, at least for the smoothing scale 
and the density range in which we are interested here.
Both stochasticity and mapping uncertainties constitute a source 
of scatter which we account for by considering 30 independent realizations
of the mass density field, as anticipated in Section~\ref{sec:lightmapping}.       

\subsection{The dark matter density field}
\label{srrtest}

As we have outlined,  our procedure of  mapping $\delms$ into $\delmu$  
enforces the correct PDF 
and power spectrum. However, the spatial correlation properties 
of the reconstructed field, $\delmsrr$, can differ significantly
from those of the true field $\delmu$.
Here we use the hydro-dynamical simulation
to estimate the errors introduced by our reconstruction
procedure   and see whether they can affect our ability of reproducing
the properties of the warm hot gas and X-ray absorbers.

We start from the original mass field 
extracted from the hydro-dynamical simulation, $\delmu$,
from which we obtain a field, $\delms$, smoothed with a 
Top Hat kernel of radius
$r_S = 200  \kms$.
Then we apply our SRR reconstruction procedure to obtain a field,
$\delmsrr$, which we compare to $\delmu$.
Finally, we smooth $\delmsrr$ with the same filter and 
compare it with $\delms$. The rationale behind this second comparison 
is that the constraints imposed on sub-resolution scales
can spoil the properties  of the reconstructed mass density
field on larger scales.

The left panel of Figure \ref{fig3} shows the one point PDF
for the original $\delta_{\rm LN}$
field in the simulation (continuous line).
The lognormal $\delmsrr$ field obtained with our
procedure (open squares) has, by construction, the same PDF as the
original one. However, on the scale of smoothing, the PDF of the 
SRR-reconstructed field
(filled triangles) is different from that of  the original, smoothed field (dashed line).
The discrepancy is significant in low density regions and vanishes 
in dense environments with $\delms \ge 10$.

The power spectra of the original and SRR-reconstructed fields have also been computed 
and are shown
in the right panel of Figure \ref{fig3} for both the smoothed and unsmoothed cases.
Also in this case, significant differences 
between the original and 
the reconstructed density fields only exits for $k>2\,\hmmpc$ i.e. on scales smaller 
than a few  Mpc. Fluctuations on a scale $r$ mostly come from waves of
wavelength $4r$, this mean that fluctuations on a scale $\ge 0.8$
Mpc are not much affected with this technique.

 We stress here that we do not aim at reproducing the power
spectrum of the simulated flux perfectly. Our reconstruction technique
is not designed to reproduce the two-point correlation properties of
the absorbers that, instead, may be significantly affected by the 
large scatter introduced by the Montecarlo sampling described i
Section~\ref{sec:igm}. Nevertheless the comparison 
with hydro-dynamical simulations shows that discrepancies  
exists at an acceptable level, even in the 2 points statistics.

Based on these results, we claim that the errors involved in 
the SRR procedure will have small impact on WHIM
structures that have a typical size of
the order of 1 Mpc and are characterized by overdensities larger than 10. In
these ranges both the PDF (left panel of Figure \ref{fig3})
and the power spectrum (right panel) of the reconstructed mass
density field are in relatively good agreement 
with those of the original simulation.

To have a qualitative view of the structures reconstructed
with the SRR procedure we plot in Figure \ref{fig4} a slice around the most massive object found
in the hydro-simulation. In the left panel we show the original mass
density field $\delmu$ while in the right one we show our reconstructed
density field, $\delmsrr$. The results  are as expected:
the densest structures and large scale features
are reproduced correctly, while on scales smaller than
that of smoothing the $\delmsrr$ is less coherent and 
the position of the small structures is not preserved.

Motivated by the quantitative and qualitative agreement between the
densest structures in the simulation and in the reconstruction
we will explore in the next
section other statistics more closely related to WHIM modeling and
detectability.

\subsection{The gas producing the absorption and the simulated spectra}
\label{spectra}
\begin{figure}
\center
\resizebox{0.55\textwidth}{!}{\includegraphics{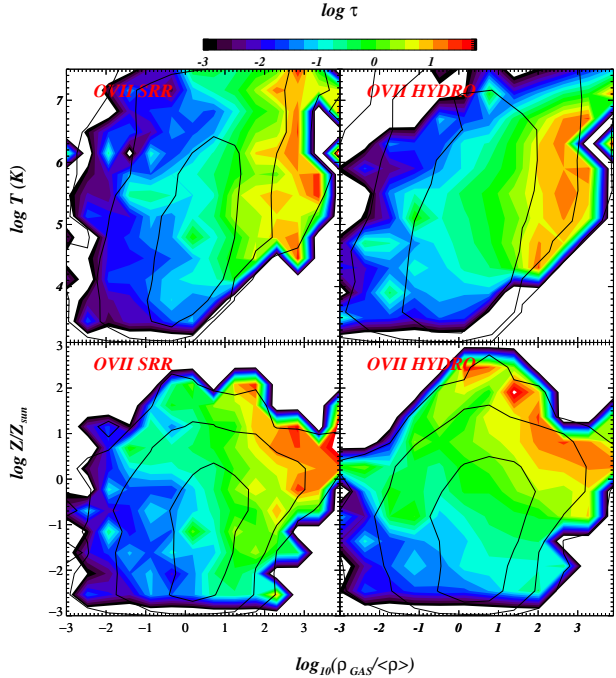}} 
\caption{Filled contour plots represent  the optical depth binned as a function of gas
density (x-axis) and metallicity (bottom panels, y-axis) and
temperature (top panels, y-axis). The results shown in the left
panels have been obtained with our reconstruction method (SRR), while the right panels 
refer to gas in the hydro-dynamical simulation (HYDRO). Empty contour plots 
(black contours) represent the
number density of pixels. The number density increases by an order of
magnitude with each contour level.}
\label{fig5}
\end{figure}

\begin{figure*}
\center
\resizebox{1.0\textwidth}{!}{\includegraphics{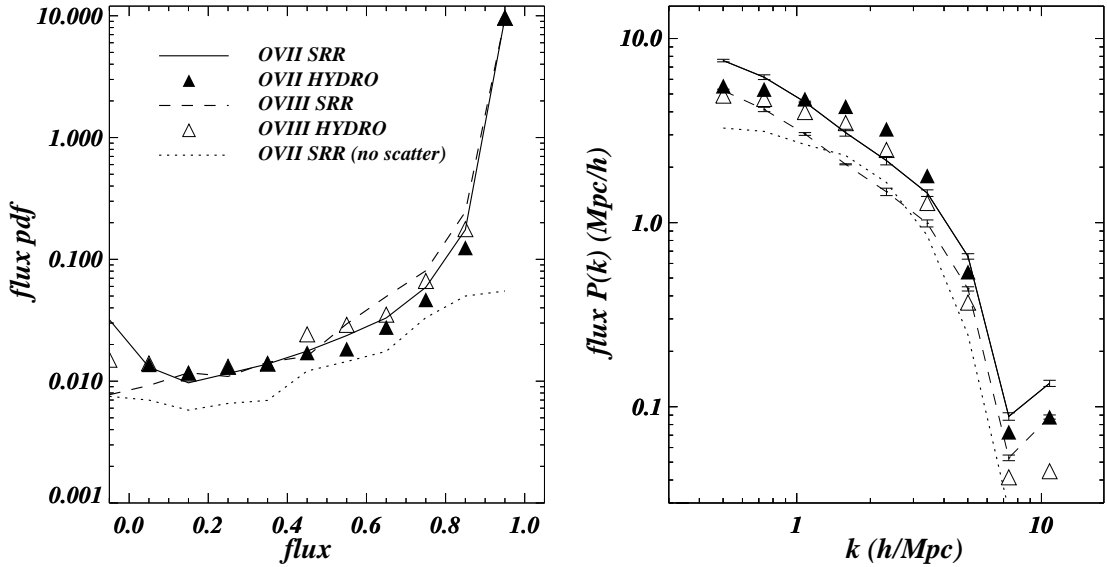}} 
\caption{{\it Left:} Probability distribution function (pdf)
for the transmitted flux in OVII (filled triangles) 
and in OVIII (empty triangles) for the hydro-simulation.
The continuous and dashed lines show the 
probability distribution function for the
reconstructed flux in OVII and OVIII, respectively. The dotted line
refers to the probability distribution function for OVII in the case in
which scatter in the reconstruction  method has been switched off.
{\it Right:} The power spectrum of the transmitted flux
in OVII and OVIII for the hydro-simulation  and 
from our reconstruction procedure. The same symbols/line-styles
are used.}
\label{fig6}
\end{figure*}

The aim of this subsection is to compare the X-ray absorption features
associated to the OVII and OVIII ions in the original
hydro-dynamical simulation with those obtained
from the full reconstruction procedure described in Section 4.
In order to do that, we draw spectra from the simulation box 
from which we compute both the optical depth and transmitted flux due to
OVII and OVIII absorbers and compare them with the 
same quantities obtained from the results of 30 independent reconstructions. 

In Figure \ref{fig5} the filled contour plots represent the optical 
depth in OVII binned as a function of the gas density and metallicity 
(bottom panels) and gas density and temperature (top panels). 
The left panels show the results of one of our
reconstructed maps (SRR) while the results in the right panel refer 
to the original hydro-dynamical simulation. 
The empty contours (black line) represent the number density
distribution of the pixels in the whole simulations (i.e. essentially
the filled contours of Figure \ref{fig1}). We have explicitly checked
that we obtain very similar results for OVIII absorbers.

This  figure shows that 
there is a reasonable agreement between the
hydro-dynamical
simulation and our reconstructions method at least for the 
optical depth of the simulated  OVII and OVIII absorbers.
This agreement indicates that our reconstruction method 
succeeds in placing the  gas responsible for
absorption in OVII and OVIII in the same environments as in the
original hydro-simulation, i.e. within regions of quite high metallicity,
$Z>0.1\,Z\odot$, and  temperature in the range $10^5<T(K)<10^{7.5}$.

In the left panel of Figure \ref{fig6} we show the
one point probability distribution function  (PDF)
of the transmitted flux  in OVII and OVIII, while in the right panel we
show the power spectrum  of the transmitted flux.
In both plots the triangles refer to the hydro-dynamical simulation while 
the continuous lines show the average over 30 independent reconstructions
performed with our technique.
The results, which are similar for both the OVII and OVIII  absorbers,
show that while the flux probability distribution functions are 
reproduced very well, as we expect 
given the constraints on the mass PDF,
the power spectra of the transmitted flux show some disagreement.
In the range $1<k\; (\hmmpc) \; <5$  the flux power spectrum of the hydro-dynamical
simulations is about $\sim 60\%$ higher than the reconstructed one. 
We note that this discrepancy is somewhat larger than the corresponding 
one for the mass density field shown in the right panel of Figure \ref{fig3}.
Moreover, in  that case  the reconstructed power at similar wavenumbers 
was systematically larger, not smaller, than the true one.
We regard this lack of power in the transmitted flux 
as determined by the large amount of scatter
involved in our modelling, that might reduce the spatial correlation 
between the absorbers on scales between 1 and 5 $\hmmpc$.
On the other hand, at wavenumbers larger than $k=0.3\; \hmmpc$, i.e.
on small scales, the power is reasonably well preserved in both the flux and the 
mass density  field.

 In addition, we show the flux probability distribution function
and the flux power spectrum for a case in which any form of scatter has
been switched off (i.e. instead of Montecarlo sampling the probability 
distribution functions as described in 
Section~\ref{sec:igm}, we simply adopt the deterministic mean relations).
We can see, as expected, that the
flux 1-point pdf is very different especially for large absorbers. This
means that the strongest absorptions in the spectra
can only be reproduced by taking into account  the large
scatter in the temperature-density and
metallicity-density relations. This confirms, at least qualitatively, the results of Chen et
al. 2003 and Viel et al. 2003. 
Switching off the scatter also affects the predicted flux of the absorbers 
that systematically underestimates that measured in the hydrodynamical simulation
by a factor $\sim 2.5$.

\begin{figure}
\center
\resizebox{0.5\textwidth}{!}{\includegraphics{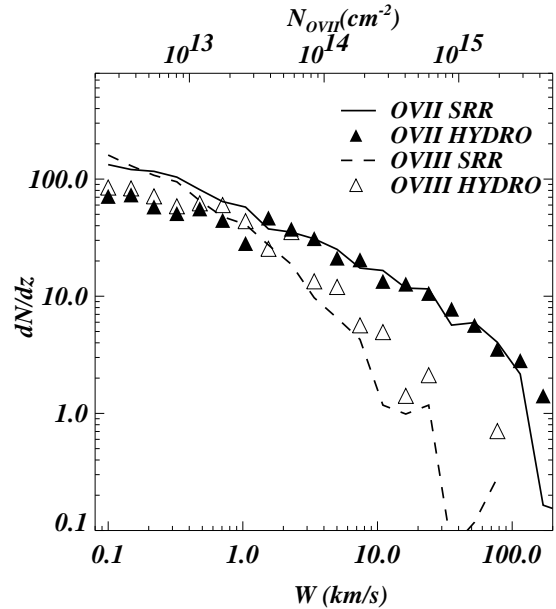}} 
\caption{The cumulative column density distribution function 
for OVII and OVII (i.e. the number of
systems per unit redshift with an equivalent width larger than a given value), as
obtained from the hydro-dynamical simulations (triangles - HYDRO) and with our
reconstruction method (lines - SRR). The symbols/line-styles are the same 
as in Figure \ref{fig6}.
The top axis shows the corresponding column density  in OVII.} 
\label{fig7}
\end{figure}

In Figure \ref{fig7} we show the cumulative number of absorbers 
in the hydro simulations, per
unit redshift range as a function of their column density (bottom
axis), i.e. the number of systems with column density larger than a
given value, and as a function of the corresponding equivalent width
(top axis), for OVII (filled triangles) and OVIII (open triangles).
The relation between 
the absorber equivalent width and the column density
is taken from Sarazin (1989)  and represents a good approximation 
for column densities smaller than $10^{15} \rm{cm}^{-2}$.
The continuous and dashed lines represent the results obtained 
from our reconstruction method and refer to OVII and OVIII, respectively. 
The comparison with the hydro-dynamical simulation
 shows that our reconstruction method is capable of
reproducing the correct number of systems with column density
$\ge 3\times 10^{13} \rm{cm}^{-2}$. Discrepancies exist
for small absorption systems that, however, could be detected neither with present
nor with next generation instruments (Viel \etal 2003).

\begin{figure}
\center
\resizebox{0.5\textwidth}{!}{\includegraphics{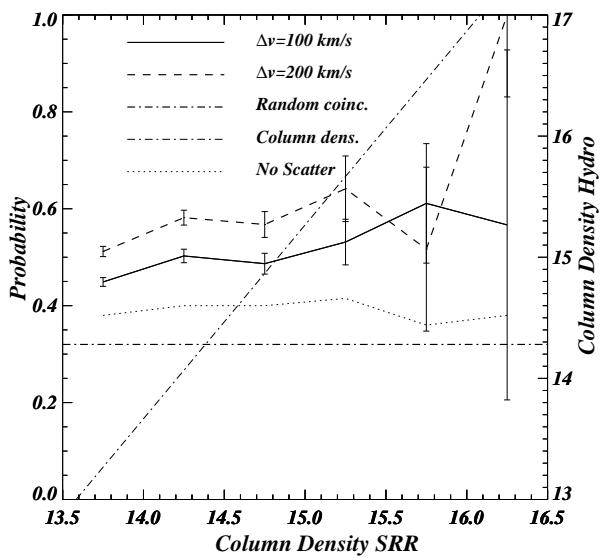}} 
\caption{
Constrained probability of finding an absorber in the hydro-dynamical simulation
map given the  detection of a corresponding absorber in the
reconstructed (SRR) maps.
This probability has been computed via hits-and-misses
statistics using 2 velocity bins of $100 \kms$ (continuous line) and of
$200 \kms$ (dashed line).  The dotted line refers to SRR simulations in
which any form of scatter has been taken out from the modelling.
Error bars represent the scatter around the mean value of the 
probability in each column density bin.
The random coincidence level for a $\Delta v
=200$ km/s is shown as the horizontal thick dot-dashed line. The thin
dot-dashed line shows the relation
between the column densities of the absorbers
reconstructed with the technique described in Section \ref{method},
and the absorbers found in the hydro-dynamical simulation
(which has to be read on the right Y-axis), for coincident systems.}
\label{fig8}
\end{figure}

As a final test, we have quantified the ability of reproducing
the correct number and equivalent width of the absorption
lines in simulated X-ray spectra by means of 
a hits-and-misses statistics.
For this purpose we have only considered the spatial distribution of 
the OVII absorbers in the hydro-simulation and in the 30 independent
reconstructions.
We use only the OVII absorbers for the reason that, for a fixed column
density, the corresponding 
equivalent width in OVII is about a factor 2 higher than the OVIII
one. This means that probably OVII absorbers can be detected in slightly
lower density environments than the OVIII ones. 

We have drawn 1000 
independent spectra in the original hydro-simulation box
and along the corresponding line of sights in each of the 
30 SRR 
reconstructions.
A positive coincidence is found when an absorber in the 
hydro simulation is identified within a bin $\Delta v= 100 \kms$
(or  $\Delta v= 200 \kms$) measured along the spectrum, 
centered on each absorber identified in  the reconstructed maps.
This procedure allows us to evaluate the constrained 
probability of finding an absorption feature in the hydro-dynamical
simulation associated to a simulated (SRR) absorber
in the reconstructed spectrum, 
binned as a function of the column density of the latter.

The results are displayed in Figure 
\ref{fig8} where it is shown that the probability an OVII
absorption line in the reconstructed spectrum
associated to an absorber identified in the hydro-simulations, ranges between
0.4 and 0.6 for the smaller velocity bin and is weakly dependent on the
column density.
For the larger velocity bin, $\Delta v = 200 \kms$, the probability is
larger and ranges from 0.5 to 1 with a stronger dependence on column
density. In this second case, all the SRR absorbers
with column density larger than $10^{16}$ cm$^{-2}$ have a
corresponding absorber in the hydro-dynamical simulations.

We note that the result is both robust, since the probability changes
by a factor $\sim 1.5$, for column densities $\mincir 10^{15.3}$
cm$^{-2}$, when using two different velocity bins $\Delta v
=100 $ and $200 \kms$ and significant,
since the error bars  are small enough for the measured 
probability to be higher that 
the random coincidence level
(thick dot-dashed line). 
In this figure we also plot the probability of finding an absorber
obtained after having switched off
any form of scatter in the temperature-density and
metallicity density relations (dotted line).
We can notice that the probability
significantly decrease  by a factor 1.3.

However, 
the fact that our method is capable of reproducing absorption features 
at the correct location $\ge 50 \%$ of the times does not necessarily imply 
that the correct column density of the absorbers is reproduced too.

The thin dot-dashed 
in Fig. \ref{fig8} shows a fit to the relation between the column density  
of the OVII absorbers 
in the reconstructions (X axis) and  the average column density
of the corresponding absorbers in the hydro-simulation  
(Y axis on the right).
The correlation  shows that on average the column density of 
the corresponding system
found in the hydro-simulations
is $\sim$ 20\% larger that the simulated (reconstructed) one.
This means that our reconstruction method tends to underestimate the
actual detectability of the WHIM absorbers.

Overall our results indicate that our reconstruction method is
capable of reproducing the spatial distribution and the physical
properties of the gas responsible for the absorptions in OVII and OVIII
and that our maps of the WHIM can be used to predict the
location and detectability of these structures in the Local universe with 
a success rate of at least $\sim 50\%$, increasing with the column density 
of the absorber.

\section{Temperature and metallicity maps of the local Universe}
\label{local}

In this Section we apply our method to investigate the spatial distribution
and the physical properties of the WHIM in the real universe.
For this purpose we apply our reconstruction procedure described in
Section 4 to the Tully catalog of nearby galaxies and produce 30
independent maps of the intergalactic gas in the local universe
that account for both the stochasticity on the biasing relation and
the scatter in the relations shown in Fig.~\ref{fig1}.
As we have explicitly checked, switching off the stochasticity
of the bias, i.e. assuming that the bias is deterministic, does not 
modify our results in term of WHIM properties (WHIM properties are in very good agreement with the maps
with stochasticity). This
means that all the sources of scatter that we add into the model, after
having produced the dark matter density field, i.e. the gas density,
gas temperature, and gas-metallicity scatter, are dominant compared to
the scatter induced by the stochasticity of the bias.

We will focus on the general properties of the local WHIM and
on our ability to map the OVII and OVIII absorbers in the local
observed universe.
In addition, we will present an example of the modelling applied to
the Virgo Cluster region in Section \ref{virgo}.

\begin{figure*}
\center
\resizebox{1.0\textwidth}{!}{\includegraphics{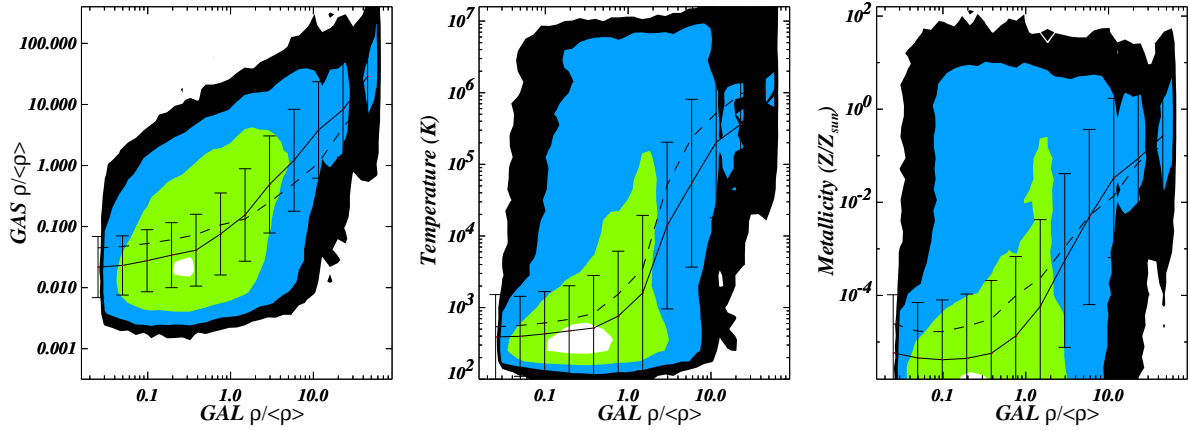}}
\caption{{\it Left:} Relation between galaxy density as derived from the
galaxy catalog and the overdensity (left panel), temperature (middle)
and metallicity (right) of the gas. The median values in each bin are shown as
continuous lines while the error bars $1\sigma$ rms value. The filled contour plots 
have the same meaning
as in Figure \ref{fig1}. The dashed line represent the same information
as extracted directly from our fiducial hydro-dynamical simulations 
(see Section \ref{hydro}.)}
\label{fig9}
\end{figure*}

The three plots of Figure \ref{fig9}, show the predicted relation
between our input galaxy luminosity field and the properties of 
our WHIM model, i.e. they illustrate to what extent the galaxy
luminosity traces the properties of the intergalactic gas 
in the local universe.
The results are very similar to those displayed in 
Figure \ref{fig1}, which, however, constitute an ingredient 
of our reconstruction scheme.
The curves and the scatter are fairly similar 
in the two cases, showing that regions of high galactic
luminosity $\delta_{\rm L}^{\rm S} \magcir 10$ are typically 
associated to the WHIM structures, although the scatter in the
temperature and metallicity of the gas is very large. In the same
Figure \ref{fig9} we overplot as a dashed line the same 
quantities as extracted from the
fiducial hydro-dynamical simulations (the 25$/h$ comoving Mpc model
$768^3$ cells rebinned into a $256^3$). We note that there is in general
a reasonably good agreement between the SRR method and the
hydro-dynamical simulations, showing once more the goodness of 
our reconstruction procedure.

\begin{figure*}
\resizebox{1.0\textwidth}{!}{\includegraphics{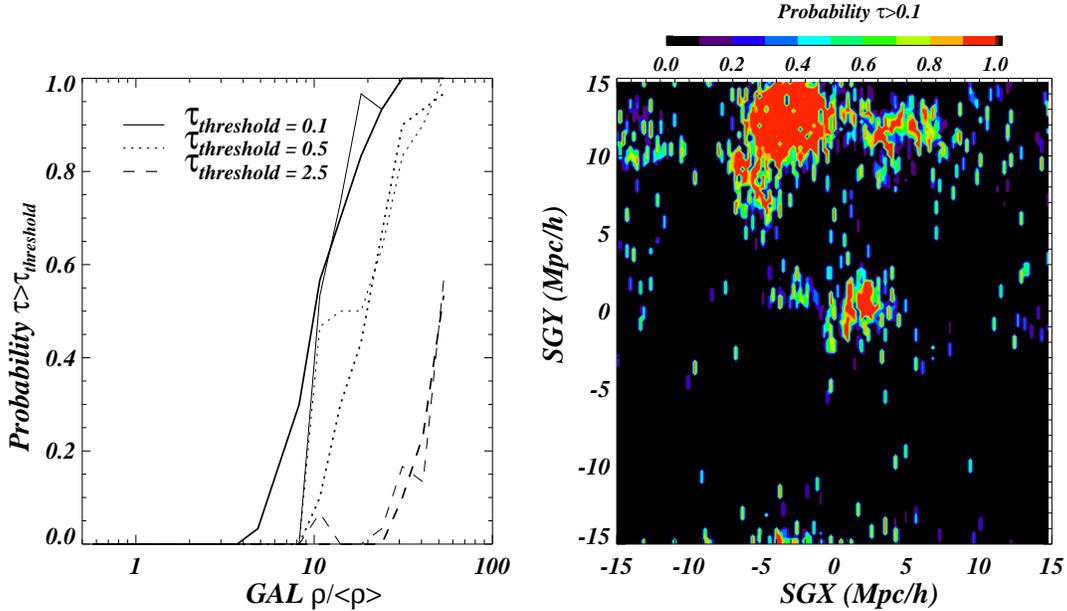}}
\caption{{\it Left:} Probability of finding an optical depth in OVII
larger than a given value, $\tau_{threshold}$, in the local universe as a function of the
density in galaxies. The thin lines indicate the same probability as
extracted in a smaller spherical region of radius $\sim 2.5 \hmpc$ around the
center of the Virgo Cluster. {\it Right:} Contour plots of the
probability of detecting an absorption with a $\tau>0.1$ in a slice of
$\sim 0.3 \hmpc$ centered around the Virgo Cluster.}
\label{fig10}
\end{figure*}

In Figure \ref{fig10} we plot the (unconstrained) probability of finding a pixel with
an OVII optical depth above a given value ($\tau_{threshold}$), calculated from the
set of 30 realizations of the local universe, as a function of the 
galaxy luminosity density in the Tully catalog. We stress that in this case we do not look for
coherent systems in the simulated maps of the universe, as we did in Section
\ref{spectra}, but we simply check for a pixel in the simulated maps to
have an optical depth in OVII above a given threshold. 
Indeed, if we assume the optical depth to be small ($\tau << 1$) and constant
within the system, then an approximate relation between equivalent width
and optical depth is given by: $W (\kms) = \tau \,\Delta v$, where $\Delta v$ is
the spatial extent of the absorber expressed in $\kms$. 
In the following we will not address the
distribution of equivalent widths (or column densities) in coherent systems. 
Rather, we keep $\Delta v$ fixed and equal to the resolution of the simulated spectra,
i.e. we consider the distribution of the optical depth in each pixel.
The rationale behind this statistics is to investigate the relation between WHIM and
overdensity in galaxies, and to quantify how likely absorption features can be detected
as a function of the density of the environment.

The left panel of Figure \ref{fig10} shows the cumulative probability
of a pixel with $\tau > 0.1$ as a function of the galaxy luminosity
density in the reconstructed maps (thick, continuous line).  Pixels
with $\tau > 0.1$ have a probability to be detected larger than 50\% in
environments with $\delta > 10$ and this probability becomes 
equal to
100\% for
$\delta \sim 30$. This means that in our model the very dense regions
are always polluted with metals that produce OVII absorptions. However,
such a small optical depth will be difficult to observe given the
uncertainties in the source continuum and the low signal-to-noise of
the observed spectra, unless they are organized in a spatially coherent
system along the line of sight.  If we ask for stronger absorptions to
take place one can see that pixels with an OVII optical depth above 0.5
(2.5) are associated with galaxy overdensities of $\delta > 16$
($\delta > 48$) with a probability larger than 50\% (35\%). In the
right panel of Figure \ref{fig10} we plot the probability map of an
optical depth larger than 0.1 in a slice of thickness $\sim 0.3 \hmpc$
containing the Virgo cluster.  The region around Virgo is visible as a
prominent clump at $(SGX,SGY)=(-3,12)$, characterized by the presence
(probability = 1) of OVII absorbers.  The thin lines plotted in the
left panel of Figure \ref{fig10} shows the cumulative probability of a
pixel with $\tau > \tau_{threshold}$ in a spherical region of radius
$2.5 \hmpc$ centered around the Virgo Cluster. In this case the
probabilities are larger than in the whole simulated local universe (up
to a factor 2 for $\tau_{threshold}=0.5$) because the coherent
structure of the Virgo cluster is quite well reproduced by our
simulated maps.

\subsection{Warm Hot Intergalactic Medium in the Virgo Cluster: a
semi-analytical study}
\label{virgo}

In the previous section we have shown that a semi-quantitative study of the
WHIM detectability in the local universe can be done with our
technique.

Motivated by the fact that WHIM structures could be detected with a
significant probability around the Virgo cluster, we study
in a more quantitative way, the 
properties of the WHIM and its detectability
in this region using our semi-analytical technique.
In addition, we investigate the robustness of our predictions
by exploring the effect of galactic winds and 
of the scatter in the relations shown in Fig.~\ref{fig1}
implemented into the model. 

\begin{figure*}
\center
\resizebox{1.\textwidth}{!}{\includegraphics{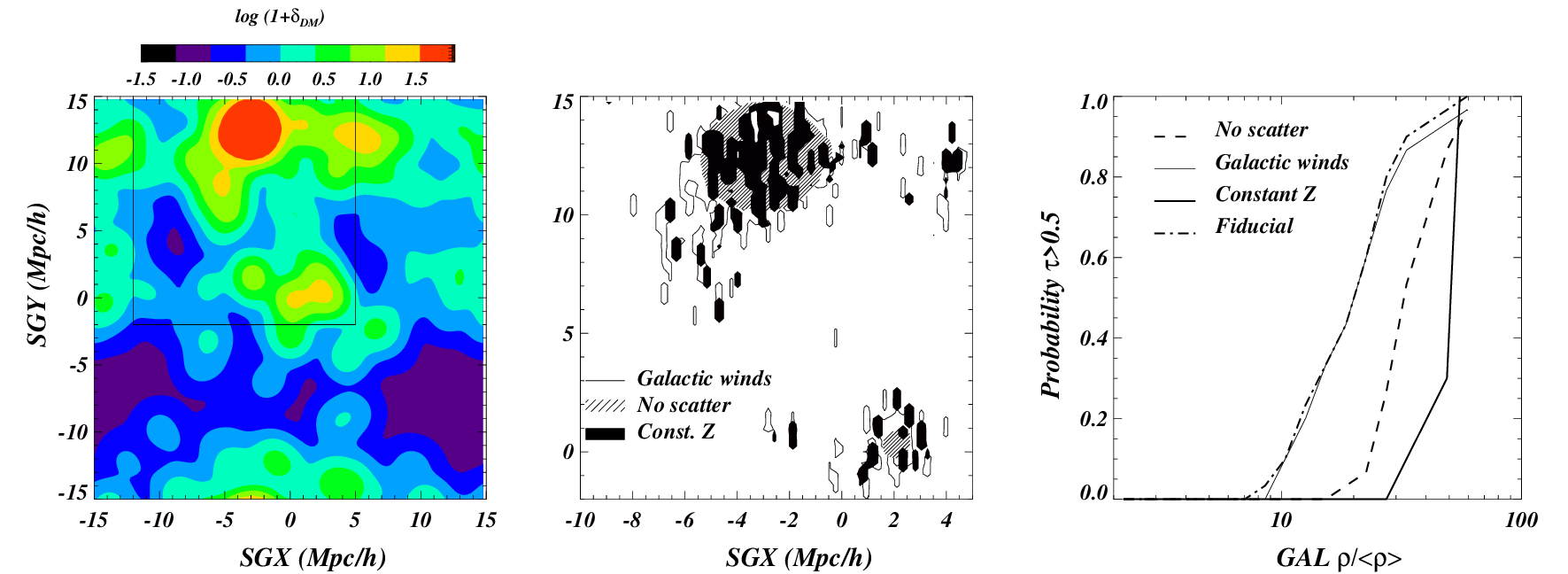}}
\caption{{\it Left:} Slice of $\sim 0.3$ Mpc$/h$ around the Virgo
Cluster representing the smoothed dark matter density field derived from the galaxy
catalog of Tully (1988). The box indicates the region around the Virgo
cluster which is considered in the other two panels. {\it Center:}
Map of the regions with probability larger than 0.5 of 
an absorber with an optical depth in OVII higher than 0.5
in the area around the Virgo Cluster. Iso-probability contours traced with
a continuous line refer to the case of strong galactic winds, with the  
metallicity-density relation right panel of Fig.~\ref{fig2}. 
The dashed regions refer to a model in which any form of scatter in the gas
density, temperature and metallicity vs. dark matter density relations
is switched off. The filled contours refer to a model in which the
metallicity has been set to a constant value of 0.3 Z$_{\odot}$, which
is the mean metallicity of the simulated box. {\it
Right:} Probability of having an absorption feature with optical depth in OVII
higher than 0.5 as a function  of galactic overdensities 
in the regions around the Virgo Cluster. 
The result of our fiducial model is represented by the dot-dashed line.
The thin continuous line represents the case of a the model with galactic winds.
The thick continuous line and the dashed line show the models with constant metallicity
and no scatter, respectively.} 
\label{fig11}
\end{figure*}

In the left panel of  Figure \ref{fig11} we show the mass density field,
obtained from the galaxy density field, smoothed
with a Gaussian filter of radius $95 \kms$, assuming the mean biasing relation
(i.e. stochasticity off).
As we have just pointed out, the Virgo
Cluster region could be very promising for detecting the WHIM component
since the typical overdensity reconstructed with our technique 
could be larger than $100$ in a region of about $(2
{\rm{Mpc}}/h)^3$ around the cluster centre, while it drops 
significantly at larger distances.

In the middle panel of Figure \ref{fig11} we zoom on a smaller region
around the Virgo cluster and we check how different implementations of
our semi-analytical modelling will affect the WHIM detectability. 
All contours represented here enclose regions with 
probability $>0.5$ of finding
an OVII absorber (pixel) with optical depth $\tau>0.5$.
In particular, since in Section \ref{local}, 
we have already explored the probability of
detecting WHIM features with our fiducial model (25$/h$ Mpc $768^3$
mesh rebinned into a $256^3$ mesh -- see Section \ref{hydro}) based on 
the relations shown in Figure
\ref{fig1}, we want to explore the effect of modifying 
those relations on the predicted  WHIM property in the 
regions around Virgo. 
We  focus on the effect of galactic winds, of the scatter in 
the different relations and on the shape of the metallicity-density relation.
The iso-probability contours enclosed by a continuous line refer to a model
in which the effect of strong galactic winds is taken into account, i.e.
when we use the metallicity-density
relation shown in the left panel of Figure\ref{fig2}, rather than that 
of Figure \ref{fig1} (while we keep the other relations to be those of
Figure \ref{fig1}).

The shaded areas
refer instead to a model in which any form of scatter in the
different relations of Figure \ref{fig1} is switched off. Finally, the filled
contours refer to a model in which the metallicity is assumed to be
constant to a value of 0.3 $Z_{\odot}$, which is the mean metallicity
of our fiducial simulation.
One can see that galactic winds, which are effective in polluting the
intergalactic medium, produce a non negligible detectability of WHIM
structures at smaller overdensities than in the other two cases. The
fiducial model, i.e. that for which the relations of Figure \ref{fig1}
hold, produces results very close to the galactic winds case.
In the model without scatter, instead, the distribution of the OVII
absorbers closely traces that of the mass, as expected.
In this case the regions containing OVII absorbers wit $\tau>0.5$
are smaller than in the
galactic wind case, meaning that the intrinsic scatter in the different
relations is effective in enhancing the WHIM detectability as found by
Chen \etal (2003) and Viel \etal (2003).
The model with a constant metallicity set to a value of 0.3
$Z_{\odot}$ tends to produce detectable regions only in very dense
environments. 
This means that a significant fraction of the absorption can come from
regions at higher metallicities than these (see Figures \ref{fig1} and \ref{fig2}).

In order to be more quantitative we plot in the right panel of Figure
\ref{fig11} the cumulative probability of a pixel with an OVII
optical depth $> 0.5$, as a function of the
galactic overdensity of the Tully catalog for the various models explored. 
The probability associated to the model with galactic winds 
(thin continuous line) is significantly larger than 
for the two other models, even at moderate overdensities of 10. 
We note however that there is very little
difference between this line and the dot-dashed one, which refers 
to our fiducial simulation, i.e. the one in which we have 
implemented the relations shown in Figure \ref{fig1}. 
It seems that winds are effective in transporting metals into the low density
regions but they produce little effects on the denser
structures in which WHIM absorption takes place. 
This means that our prediction for the WHIM distribution are robust
since they are weakly affected by the presence  
of strong galactic winds. This issue will be
investigated in more detail in a future work.
In the rather unphysical case of pure deterministic relations
between the IGM properties, i.e. the case of no scatter
(dashed line),
the probability decreases by a factor of 2
in the range of overdensities 20-40.
The lack of any correlation between density and metallicity 
(thick continuous line) has an even more dramatic effect
since in this case one could  hope to detect OVII 
absorptions features only in the few  higher density regions
(although this of course depends on the 
value chosen for the constant metallicity level).

We stress here that the modelling presented in this paper offers not
only predictions for observers on where to look at in order to detect WHIM
structures but allows to put constraints on the state of the IGM even
if OVII and OVIII lines are not detected along a particular line-of-sight.

\section{Discussion and conclusions}

Motivated by the recent interest in the warm-hot phase of the intergalactic medium 
and  by the present difficulty of its detection, we have developed a semi-analytic model
capable of predicting the spatial distribution of the WHIM in the local supercluster,
its physical properties and the probability of detecting it through absorption lines
in the X-ray spectra of bright background extragalactic sources.

Unlike the models previously proposed by Kravtsov, Klypin \& Hoffman (2002)
and by Yoshikawa {\it et al.} (2004) our technique does not require
to perform a full hydro-dynamical simulation.
Instead, since it uses the already existing outputs of numerical simulations,
it requires much less CPU time to model the WHIM in the local 
universe from the observed high resolution map of galaxy light density.
As a consequence we are able to
{\it i)}  trace the distribution of mass and 
gas with a  resolution of $\sim 1  \hmpc$ (Gaussian) and 
{\it ii)} perform several
independent realizations to account for the model uncertainties, through which
we can quantify the probability of detecting an absorption feature at a given 
redshift in an X-ray spectrum drawn along an arbitrary direction in the sky.

The most serious drawback of our technique is the lack of self-consistency 
in the reconstruction procedure which results in a lack of spatial coherence
on sub-Mpc scales that prevents us from reconstructing the
properties of the IGM in the local universe on a point-by-point basis.
However, as we have explicitly checked through extensive numerical tests, our 
model  correctly predicts the spatial location and the physical conditions
of the WHIM, that typically resides in regions of enhanced densities.
We have focused on the OVII and OVIII ions that are responsible
for the strongest WHIM absorption features in the X-ray spectra and have shown that 
our model correctly reproduces the statistical properties of the WHIM absorbers.
In particular, we are able to reproduce the correct 
number of absorption systems 
per unit redshift as a function of the column density,
especially for absorbers with large equivalent width.
Moreover, we are able to predict the correct spatial location of the OVII
absorbers 40-70\% of the times, significantly above the random level,
and up to 100\% for  systems with column density of $\sim 10^{16} {\rm cm}^{-2}$, 
although we somewhat underestimate their actual equivalent width.
Finally, our semi-analytic technique is capable of reproducing the global physical 
properties of the IGM, quantified in terms of gas density, temperature and metallicity 
vs. galaxy light density relations.

It is worth pointing out that several alternative routes to our reconstruction 
procedure are also possible. A very interesting one, proposed by 
the anonymous referee, consists of mapping galaxy light density to OVII and OVIII density 
without any reference to the matter density field. The idea is to determine 
the distribution of OVII and OVIII in the hydrodynamical simulation and use their 1-point PDF
in the inversion procedure described in Section 4.1 to map galaxy light to  OVII and OVIII 
density in the local universe, hence avoiding  
mapping mass to intergalactic gas (Section 4.3) which represents the main source of scatter 
in our method.
Unfortunately, as we have verified, the mean OVII and OVIII vs galaxy density relation
flattens out and becomes non monotonic when the density increases. 
As a result, it is not possible to trace OVII and OVIII in regions of enhanced densities where 
the stronger absorbers typically resides unless the inversion method of Section 4.1 
could be suitably modified.

Encouraged by our ability of mapping the distributions of OVII and OVIII
in our local universe, we have focused on the region around the Virgo Cluster.
We found a large ($>0.5$) probability of an OVII absorber with opacity ($\tau >0.5$) in 
each resolution element contained in a quasi-spherical region of radius $2.5 \hmpc$ centered
on the Virgo Cluster that therefore represents the most promising site for detecting WHIM
in the local universe (hence confirming, at least qualitatively,
the results of Kravtsov, Klypin \& Hoffman (2002)).
This prediction is robust, since the size of this region and the probability detection level do not 
change significantly when one allows for uncertainties in  modeling the IGM properties.
In addition we have studied the effect of strong galactic winds that are effective
in polluting low density regions, but seems not to 
appreciably affect regions of enhanced density where
OVII and OVIII absorbers are preferentially located.
We have found that using deterministic relations to describe the IGM properties 
can be quite dangerous, as it would result in a significant underestimate of
the occurrences of OVII and OVIII absorbers 
in low density regions (an effect that becomes even more dramatic when 
assuming an IGM with constant metallicity).

This semi-analytic model is therefore able to map the probability of
detecting absorbers associated to the highly ionized ions in the WHIM,
and to predict the equivalent width of the corresponding absorption
features in X-ray spectra.  This is of considerable observational
interest since it allows us to quantify the probability of observing
absorption features in the spectra of all bright X-ray sources,
produced by the WHIM in the Local Supercluster.  Moreover, since our
method somewhat under predicts the column density of the absorber, we
will also be able to provide a lower limit to the significance of
detecting such features with a specified instrument. It is worth
stressing that this method offers the possibility to put constraints on
the physical properties of the IGM (metallicity, temperature,
ionization state) even in the presence of non detections of OVII and
OVIII absorbers along a particular line-of-sight and not only when
these WHIM absorption features are detected.

Obviously, our model is also able predict the emission lines produced by the very ions responsible
for the X-ray absorption features very much like in the recent study of  Yoshikawa {\it et al.}
(2004), although on a much more local basis.

Our model predictions can be cross-correlated with existing X-ray
spectra to increase the significance of marginal detections of
absorption lines.

It is worth stressing that while our modeling is performed in real space, absorption features
are observed in redshift space. Redshift space distortions can be easily accounted for in our modeling
by displacing the line-of-sight position of the X-ray absorbers according to some 
of the available model velocity field, like the non-parametric 
one of Branchini {\it et al.} (1999). These corrections are expected to be quite minor
since the cosmic flow in the local supercluster is notoriously cold (Klypin 
{\it et al.} 2003).

Our model is currently limited to a very local volume of the universe. This 
is the consequence of having required a very dense sampling in tracing the galaxy light.
In fact, our current implementation of the light to mass tracing technique requires 
at least one object per resolution element, a requirement that sets the maximum size 
of the explored volume. More sophisticated techniques, like the one recently proposed by
Szapudy \& Pan (2004), can be successfully implemented in case of much sparser sampling and thus  
can be used to extend our WHIM predictions  beyond  the boundaries of the local supercluster.

\section*{Acknowledgements.}
We would like to thank Adi Nusser, Yehuda Hoffman and Saleem Zaroubi for very 
useful discussions and suggestions. We thank Kentaro Nagamine for
providing us with an estimate of the stochasticity in the biasing relation
from hydro-dynamical simulations.
We also wish to thank the anonymous referee for suggesting us an
original and alternative mapping procedure.
This work has been partially supported by the European Community
Research and Training Network `The Physics of the Intergalactic
Medium', by grant ASC97-40300 and by the National Science Foundation
under Grant No. PHY99-07949. MV thanks PPARC and
the Kavli Institute for Theoretical Physics in Santa Barbara for
financial support during the program ``Galaxy-Intergalactic Medium Interactions''. Part
of this work has been performed on COSMOS (SGI Altix 3700)
supercomputer at the Department of Applied Mathematics and Theoretical
Physics in Cambridge. COSMOS is a UK-CCC facility which is supported by
HEFCE and PPARC.

\end{document}